\documentstyle[aps,twocolumn]{revtex}
\begin{document}
\title{Rapidly Rotating Fermi Gases}
\author{Tin-Lun Ho and C. V. Ciobanu}
\address{Department of Physics,  The Ohio State
University, Columbus, Ohio 43210}

\maketitle

\begin{abstract}
We show that the density profile of a Fermi gas in rapidly rotating
potential will develop prominent features reflecting the underlying
Landau level like energy spectrum. Depending on the
aspect ratio of the trap, these features can be a
sequence of ellipsoidal volumes or a sequence of quantized steps.
\end{abstract}

Currently, there is an intense effort to cool trapped Fermi gases down to the
degenerate limit. Recent experiments at JILA on $^{40}$K has reached one
half of its Fermi temperature\cite{Jin}.
One of the motivations of
cooling Fermions is to reveal their possible superfluid ground states, which
can be quite novel in the case of multi-component
Fermion systems\cite{HoYip}. Current theoretical estimates, however, indicate
that the interactions between different spin states of $^{40}$K are
all positive\cite{Burke},
implying a normal instead of superfluid ground state.
The absence of superfluid ground states, however, does not mean that the
system can not have novel {\em macroscopic} quantum phenomena.
Quantum Hall effect is an excellent example. In a strong magnetic field,
the energy levels of a two
dimensional electron system organize into highly degenerate Landau levels,
leading to a whole host of dramatic effects. In this paper, we show that
similar organizations
will take place in a fast rotating three
dimensional trapped Fermi gas, leading to many macroscopic quantum phenomena.

For neutral atoms in rotating harmonic traps, we shall see that Landau level
like
energy spectrum will appear when the rotation frequency $\Omega$  approaches
the transverse confining  frequency $\omega_{\perp}$.
This ``fast rotating" limit might appear hard to achieve
as the system is at the verge of flying apart due to centrifugal
instability\cite{rotbose}.
Such instability, however, can be prevented by imposing an
additional repulsive potential which dominates over the centrifugal force
beyond certain radius.  With centrifugal instability eliminated, the
Landau-like levels will show up in many ways.  We shall see that
for cylindrical harmonic traps with $\omega_{z}<<\omega_{\perp}$, where
$\omega_{\perp}$ and $\omega_{z}$ are the transverse and longitudinal trapping
frequencies, the density is a sum of one-dimensional like
density distributions residing in different ``Landau volumes''.
For traps with $\omega_{z}$ comparable or smaller than $\omega_{\perp}$,
the density consists of a set discs along $z$, each of which is
made up of a sequence of density steps quantized in units of
$M\omega_{\perp}/(\pi \hbar)$.

{\bf 2D case}:
We first consider 2D rotating Fermi gases in harmonic
potentials since they illustrate the basic physics.
The Hamiltonian in the rotating frame is
\begin{equation}
H - \Omega L_{z} = \frac{1}{2M}{\bf p}_{\perp}^2 +
\frac{1}{2}M\omega_{\perp}^2 r^2 -
\Omega \hat{\bf z}\cdot {\bf r}\times {\bf p}_{\perp},
\label{HL} \end{equation}
where ${\bf p}_{\perp}= (p_{x}, p_{y})$ and ${\bf r}=(x,y)$.
The eigenfunctions and eigenvalues
of eq.(\ref{HL}) are
\begin{equation}
u_{n,m}(r,\theta) = \frac{e^{|w|^2/2} \partial_{+}^{m}\partial_{-}^{n}
 e^{-|w|^2} }{\sqrt{ \pi a_{\perp}^{2}n! m!}}
\label{wf} \end{equation}
\begin{equation}
\epsilon_{n,m}=\hbar (\omega_{\perp} + \Omega)n + \hbar (\omega_{\perp} -
\Omega)m + \omega_{\perp},
\label{energy} \end{equation}
where $n,m$ are non-negative integers $0,1,2,..$; $w\equiv (x+iy)/a_{\perp}$;
 $a_{\perp}= \sqrt{\hbar/M\omega_{\perp}}$ ; and
$\partial_{\pm} \equiv (a_{\perp}/2)(\partial_{x} \pm i \partial_{y})$.
To derive
eqs.(\ref{wf}) and (\ref{energy}), we note that eq.(\ref{HL})
can be written as ${\bf \Pi}^2/2M + (\omega_{\perp} - \Omega)L_{z}$ with
${\bf \Pi} = {\bf p}_{\perp} - M\omega_{\perp}\hat{\bf z}\times {\bf r}$,
which is precisely the canonical momentum ${\bf \Pi}={\bf p}_{\perp} -
\frac{eB}{2c}\hat{\bf z}\times{\bf r}$ of an electron
in a magnetic field ${\bf B}$ in the symmetric gauge,
with $eB/Mc= 2\omega_{\perp}$. The eigenfunctions of
${\bf \Pi}^2/2M$ are those in eq.(\ref{wf})\cite{Laughlin}, with
eigenvalues $\epsilon_{n,m}^{L}=\hbar\omega_{\perp}(2n+1)$,
where $n$ is the Landau level index, and $m$ is an ``angular momentum"
index labelling the degeneracy in each level. Since $L_{z}u_{n,m} =
\hbar (m-n)u_{n,m}$, eq.(\ref{wf}) is also an eigenstate
of eq.(\ref{HL}) with  eigenvalues eq.(\ref{energy}).
Note that the function $u_{n,m}$ in  eq.(\ref{wf}) peaks at
\begin{equation}
r_{n,m}\equiv \langle r^2 \rangle_{n,m} = a_{\perp}^2(n+m+1),
\label{rms} \end{equation}
and decays away as a Gaussian over a distance $a_{\perp}$.

Eq.(\ref{energy}) shows that the system is unbounded
when $\Omega>\omega_{\perp}$
unless an additional repulsive potential $V_{\rm wall}(r)$
(say, introduced by an additional optical trap) is present.
We shall in particular consider potentials $V_{\rm wall}(r)$ which are zero
for $r<R$ but become strongly repulsive for $r>R$, with $R>>a_{\perp}$.
The specific  form of $V_{\rm wall}$
is not important  for the key features  discussed below, as long as it is
smooth over length scale $a_{\perp}$.
The condition $R>>a_{\perp}$, however, allows us to fit many $m$ states
inside $r<R$ and is a necessary feature for
many effects discussed below. Since $V_{\rm wall}(r)$ is cylindrically
symmetric, the eigenstates are still labeled by quantum
numbers $(n,m)$.  For states originally with $r_{n,m}<R$,  eqs.(\ref{wf}) and
(\ref{energy}) remain valid because $V_{\rm wall}=0$ for $r<R$.
For states $(n,m)$ originally peaked beyond $R$,
their energies increase rapidly because $V_{\rm wall}$ is strongly repulsive.
(For $^{40}$K in a tight trap
$\omega_{\perp} = 4000$Hz and $10^5$Hz, we have $a_{\perp}\approx 2.5\times
10^{-5}$cm and $5\times 10^{-6}$cm resp.  The condition
$R>>a_{\perp}$ is satisfied for $R>5\times 10^{-4}$cm.)

Let us first consider the case $\Omega<\omega_{\perp}$ with $V_{\rm wall}=0$.
The density in the ground state is
$\rho(r) = \sum_{n,m}|u_{n,m}({\bf r})|^2\Theta(\mu-\epsilon_{n,m})$,
where $\Theta(x)=1$ or $0$ if $x>0$ or $<0$,
$\mu$ is the chemical potential related to the particle number $N$ as
$N= \sum_{n,m}\Theta(\mu-\epsilon_{n,m})$. We can write
$\rho(r) = \sum_{n=0}^{n^{\ast}}\rho_{n}(r; m^{\ast}_{n})$,
where $\rho_{n}(r; L)$ is density contribution of the $n$-th Landau level
with  angular momentum states filled up to  $m=L$;
\begin{equation}
\rho_{n}(r; L) = \sum_{m=0}^{L}|u_{n,m}({\bf r})|^2 \Theta(\mu -
\epsilon_{n,m}),
\label{rhonL} \end{equation}
$m^{\ast}_{n}$ is the highest angular momentum state in
the $n$-th Landau level with energy less than $\mu$, and $n^{\ast}$ is the
highest Landau level below $\mu$,
\begin{equation}
m_{n}^{\ast} = {\rm Int}\left[\frac{\mu/\hbar-\omega_{\perp}-(\omega_{\perp}+\Omega)n}{(\omega_{\perp}-\Omega)}\right],
\label{mast} \end{equation}
$n^{\ast}$$={\rm Int}$
$\left[\frac{\mu/\hbar -\Omega}{\omega_{\perp}+\Omega}\right]$,
where Int$[x]$ denotes the integer part of $x$,
and $x$ is understood to be positive. Since $(n,m^{\ast}_{n})$ is the
state in $\rho_{n}$ farthest from the origin, its peak location
($r_{n}=r_{n,m^{\ast}_{n}}=a_{\perp}\sqrt{n+m^{\ast}_{n}+1}$) gives the size
of $\rho_{n}$.
 When $\Omega$ is very close to $\omega_{\perp}$, we have
$m_{n}^{\ast}>>1$ and
\begin{equation}
r_{n}^2 =\frac{\mu - \hbar \Omega(2n+1)}{M\omega_{\perp}(\omega_{\perp} -
\Omega)}.
\label{rn} \end{equation}
Note that the difference in area between successive  Landau discs is a
constant
\begin{equation}
\pi(r_{n-1}^2 - r^{2}_{n}) = (\pi a_{\perp}^2) \left(
\frac{2\Omega}{\omega_{\perp} - \Omega} \right).
\label{rn2} \end{equation}

Using eq.(\ref{wf}), it is straightforward to show that
\begin{equation}
\rho_{0}(r; m^{\ast}_{0}) = \frac{1}{2 \pi a_{\perp}^2}
\left[ 1 - {\rm erf}\left(\frac{s}{\sqrt{2m_{0}^{\ast}}}\right)(1+ [..])
\right].
\label{rho0} \end{equation}
where $s=(r/a_{\perp})^2-m^{\ast}_{0}$,
${\rm erf}(x) = (2/\sqrt{\pi})
\int^{x}_{0}e^{-z^2}{\rm d}z$, and the term $[..]$ in
eq.(\ref{rho0}) is of order $(12m_{0}^{\ast})^{-1}$ and smaller.
The densities $\rho_{n}(r)$  of the higher Landau levels
$(n>0)$ can be generated from $\rho_{0}$ as\cite{comment}
$\rho_{n}(r; m^{\ast}_{n})$$=\frac{1}{n!}$
$(\frac{1}{4}\nabla_{\lambda}^2)^{n}$
$[\rho_{0}(\vec{r}-a_{\perp}\vec{\lambda};m^{\ast}_{n})$
$e^{\vec{\lambda}^{2}}]_{\vec{\lambda}=0}$.
>From the properties of ${\rm erf}(x)$, it is clear that
$\rho_{0}$ is a constant $1/(\pi a_{\perp}^2)$ within a disc of radius
$r_{0}=\sqrt{m_{0}^{\ast}+1} a_{\perp}$ and has an edge of
width $\Delta_{0}\approx \frac{3}{\sqrt{2}}a_{\perp}$\cite{width}.
If $m_{0}^{\ast}>>1$, $\Delta_{0} << r_{0}$,
and $\rho_{0}$ can be approximated as a step function on scales
larger than $\Delta_{0}$. Likewise, $\rho_{n}$
can be approximated as a step function with somewhat larger width
within the same approximation.
Thus, when $m^{\ast}_{n}>>1$, we have
\begin{equation}
\rho_{n}(r) = (\pi a_{\perp}^2)^{-1}\Theta(r_{n}^2 - r^2).
\label{step} \end{equation}
Eq.(\ref{step}) and (\ref{rn2}) then imply
$\rho(r) = \frac{M\omega_{\perp}}{\pi \hbar}I(r)$, with
$I(r) = \sum_{n}\Theta \left(
\mu - M\omega_{\perp}(\omega_{\perp} - \Omega)r^2
- \hbar\Omega(2n+1)  \right)$.
Using the identity
${\rm Int}[x+1] = \sum_{n} \Theta(\alpha (x-n))$, for all $x>0$ and $\alpha>0$,
$\rho(r)$ can be simplified to
\begin{equation}
\rho(r) = \frac{M\omega_{\perp}}{\pi \hbar} {\rm Int} \left[
\frac{\mu - M\omega_{\perp}(\omega_{\perp} - \Omega)r^2 + \hbar\Omega}
{2\hbar\Omega} \right].
\label{fir} \end{equation}

It is, however, instructive to re-derive eq.(\ref{fir})
in a way generalizable to arbitrary potentials.
We rewrite eq.(\ref{HL}) as
\begin{equation}
H - \Omega L_{z} = \frac{({\bf p}_{\perp} - M\Omega \hat{\bf z}\times {\bf
r})^2}{2M} +  \frac{1}{2}M(\omega_{\perp}^2 - \Omega^2)r^2 .
\label{2DH} \end{equation}
The first term gives a set of Landau levels with spacing
$2\hbar \Omega$, each of
which contributes $(\pi A^{2})^{-1}$ to the density,
where $A^{2} = \hbar/(M\Omega)$.
If the second term in eq.(\ref{2DH}) were absent, the density is given by
$\rho = I/(\pi A^{2})$, where $I$ is the total number of Landau levels below
the chemical potential, $I={\rm Int}[\frac{\mu+\hbar \Omega}{2\hbar\Omega}]$.
When $\Omega$ is close to $\omega_{\perp}$,
the second term in eq.(\ref{2DH}) is
slowly varying over the scale of $A\sim a_{\perp}$ and can be absorbed
in the chemical potential. The density profile within local density
approximation (LDA) is then
\begin{equation}
\rho(r) = \frac{M\Omega}{\pi \hbar}I(r), \,\,\,\,\,\,
I(r) = {\rm Int}\left[\frac{\mu(r) + \hbar \Omega}{2\hbar\Omega}\right]
\label{local} \end{equation}
where $\mu(r) = \mu - \frac{1}{2}M(\omega_{\perp}^{2}-\Omega^2)r^2$.
Clearly, eq.(\ref{local}) is equivalent to (\ref{fir})
up to correction $(1 - \Omega/\omega_{\perp})$$<<1$ as $\Omega\rightarrow
\omega_{\perp}$.
Eq.(\ref{local}), once established, is easily generalized
to other potentials. In the presence of $V_{\rm wall}$, one simply replaces
$\mu(r)$ in eq.(\ref{local}) by
\begin{equation}
\mu(r) = \mu - \frac{1}{2}M(\omega_{\perp}^2 - \Omega^{2})r^2 -
V_{\rm wall}(r).
  \,\,\,\,\,\,\,  (2D)
\label{mur} \end{equation}
Eq.(\ref{local}) and eq.(\ref{mur}) constitute the LDA solution for
the 2D rotating Fermi gas for both
$\Omega<\omega_{\perp}$ and $\Omega>\omega_{\perp}$. The schematics of
LDA is shown in fig.1(a) and 1(b).

To understand the validity of LDA (eq.(\ref{local})),
we have calculated the density numerically using eq.(\ref{wf}).
The result for a system of $2000$ Fermions at
$\Omega/\omega_{\perp}= 0.996$ is shown in figure 2a. The system
exhibits a sequence of quantized steps at locations well described by LDA.
The evolution of the density within the range $0.99<\Omega/\omega_{\perp}<1$
of this Fermion system is shown in fig.2b.
As $\Omega$ decreases, more Landau levels are populated while the steps near
the surfaces are closer together, (as expected from eq.(\ref{mast}) and
(\ref{rn2})), yet the step structures remain discernable and correctly described
by LDA, (the LDA construction is not displayed so as to keep the fig.2b
readable). The behaviors of the densities at lower frequency
$0.98<\Omega/\omega_{\perp}<1$ are shown in fig.3a and 3b for a system of 1000
Fermions. At the lowest frequency displayed, (fig.3a), the step structure near
the surface is completely smeared out by the spread of the edges. Nevertheless,
the density of the innermost plateau remained {\em quantized}, with a size
correctly described by LDA. This core of quantized density (or ``quantized
core" for short) is a clear evidence for Landau levels.
Our studies show that for about 2000 particles, Landau levels will show up as
a sequence of discernable steps only when $0.99<\Omega/\omega_{\perp}<1$,
which is achievable with the current capability to control frequencies,
especially for large $\omega_{\perp}$. For lower frequencies, the existence of
Landau levels can only be revealed through the presence of a ``quantized core",
which shrinks in size as $\Omega$ decreases.
On the other hand, the LDA in fig.1b shows that by introducing an additional
potential $V_{\rm wall}$, Landau levels (in the form of a sequence of steps
near the center or a ``quantized core") can still be revealed at frequencies
farther beyond $\omega_{\perp}$, even though the steps near the surface are
smeared out.

{\bf 3D case} : For a 3D harmonic trap, eq.(\ref{HL}) acquires terms
$p_{z}^2/2M + \frac{1}{2}M\omega_{z}z^2$, which give rise to harmonic
oscillator eigenfunctions $f_{n_{z}}(z)$ with eigenvalue
$\epsilon_{n_{z}}=\hbar\omega_{z}(n_{z}+\frac{1}{2})$.
The density is
$\rho(r,z)$$=\sum_{n_{z},n,m}$$|f_{n_{z}}(z)|^2$$|u_{n,m}({\bf r})|^2$
$\Theta(\mu-\epsilon_{n_{z},n,m})$,
where $\epsilon_{n_{z},n,m} = \epsilon_{n_{z}} + \epsilon_{n,m}$.
When $\Omega/\omega_{\perp}\sim 1$, we first perform the $m$-sum because
it produces the smoothest
change in the energy. Repeating the steps leading to eq.(\ref{step}), we have
\begin{equation}
\rho(r, z) = \sum_{n_{z}, n}\frac{|f_{n_{z}}(z)|^2}{\pi a_{\perp}^2}
\Theta(\mu(r)  - \hbar \Omega(2n +1) - \epsilon_{n_{z}})
\label{density3} \end{equation}
where $\mu(r) = \mu - M\omega_{\perp}(\omega_{\perp} - \Omega)r^2 \approx \mu -
\frac{M}{2}(\omega_{\perp}^2 - \Omega^2)r^2$.

The order of summation of the remaining integers $n_{z}$ and $n$
depends  on the relative  strength between $\omega_{z}$ and
$\omega_{\perp}$.  When $\omega_{z}<<\omega_{\perp}$, we first sum $n_{z}$.
To do that, we note that the density of a 1D Fermi gas
is $\rho_{1D}= \sqrt{2M\mu}/(\pi \hbar)$, where $\mu$ is the chemical
potential. In a harmonic trap, the density within LDA is obtained by
changing $\mu$ to $\mu(z)= \mu - \frac{1}{2}M\omega_{z}^2 z^2$. We then have
$\rho_{_{1D}}(z)=\sum_{n_{z}}|f_{n_{z}}(z)|^2 \Theta(\mu - \epsilon_{n_{z}})
=  \frac{\sqrt{2M\mu(z)}}{\pi \hbar}$, which
turns eq.(\ref{density3}) into a useful LDA
\begin{equation}
\rho(r) = \frac{1}{\pi^2} \frac{1}{a_{\perp}^2 a_{z}^{}}
\sum_{n} \sqrt{\left[\mu(r,z) - \hbar \Omega(2n+1) \right]/
(\frac{1}{2}\hbar \omega_{z})},
\label{densqrt} \end{equation}
where  $a_{z} = \sqrt{\hbar/M\omega_{z}}$,
and $\mu(r,z) = \mu - \frac{M}{2}(\omega^{2}_{\perp}- \Omega^2)r^{2} -
\frac{M}{2}\omega_{z}^2 z^2 - V_{\rm wall}$. We have included $V_{\rm wall}$
in $\mu$ for the more general situation as in the 2D case.
Eq.(\ref{densqrt}) shows that $\rho(r)$
is a sum of 1D densities (labelled by
``$n$") each of which distributed over in a ``Landau volume" bounded by
the ``Landau surface" $\mu(r,z) = \hbar \Omega (2n+1)$.
When $V_{\rm wall}=0$, $\Omega<\omega_{\perp}$
the Landau surface are ellipsoidal surfaces. It is easy to verify that
the surface areas $A_{n}$ for successive ellipsoids differ by a constant
$A_{n-1} - A_{n} = \frac{\hbar}{M\Omega}\left(
\frac{16\pi\Omega^2}{\omega_{\perp}^2 - \Omega^2}\right)$.
When $\Omega \geq \omega_{\perp}$, centrifugal instability against harmonic
confinement sets in and stability can only be established by $V_{\rm wall}$.

To demonstrate the validity of the LDA eq.(\ref{densqrt}), we have evaluated
the density numerically for a system of 2000 Fermions for
$\omega_{z}/\omega_{\perp}=0.2$
at $\Omega/\omega_{\perp}=0.99$. The results are shown in figure 4. It
shows that the LDA (dotted line)
is a good approximation.
The Landau volumes can be clearly identified
by the change of slope in the
density. The appearance of a plateau the center is because
$\omega_{z}/\omega_{\perp}$ is only 0.2, revealing the 2D feature
of the $n_{z}$ levels.
For smaller ratios of $\omega_{z}/\omega_{\perp}$, the plateau
disappears and the LDA expression (eq.(\ref{densqrt})) is achieved.

When $\omega_{z}>\omega_{\perp}$, summation of $n$ in eq.(\ref{density3})
gives
\begin{equation}
\rho(r) = \frac{M\omega_{\perp}}{\pi \hbar}
\sum_{n_{z}}|f_{n_{z}}(z)|^2 {\rm Int}\left[
\frac{ \mu(r) - \hbar \omega_{z}(n_{z} + \frac{1}{2})
+ \hbar\Omega  }{2\hbar\Omega} \right]
\label{nextdensity} \end{equation}
where
$\mu(r)= \mu - \frac{1}{2}M(\omega_{\perp}^2 - \Omega^2)r^2 - V_{\rm wall}$.
In this limit, the density consists of a sequence of discs (labeled by $n_{z}$)
in the $z$-direction. Each disc $|f_{n_{z}}(z)|^2$ consists of a sequence of
density steps in the $xy$-plane reflecting the number of filled Landau levels.
The behavior of the density within each disc in the $xy$-plane is identical to
the 2D case discussed before. Finally, we note that as temperature increases,
the Landau level structure near the surface will first melt away, and the
melting will proceed
toward the center. The temperature below which the Landau level effect
 begin to appear is  $T=2\hbar\omega_{\perp}/k_{B}$, which is $3.8\times
10^{-7}$K and $9.6\times 10^{-6}$K for
for $\omega_{\perp}=4000$Hz and $10^5$Hz resp., a temperature range
achievable in current experiments\cite{temp}.

So far, we have only discussed the effect of Landau levels
on the density profiles of fast rotating Fermi gases.
If the development of quantum Hall effect in the last decade is a guide,
one expects many more novel phenomena in Fermi gases
in the fast rotating regime.
This work is completed during a workshop at the Lorentz Center of University
of Leiden. We thank Professor Henk Stoof and the Lorentz Center for
generous support. This work is supported by a Grant
from NASA NAG8-1441, and by the NSF Grants DMR-9705295 and  DMR-9807284.

\noindent Caption

\noindent Fig.1a and 1b:
Fig.1a and 1(b) corresponds to $\Omega<\omega_{\perp}$ and
$\Omega>\omega_{\perp}$ resp. The rapid drop at large $r$ is due to
the strongly repulsive potential $V_{\rm wall}$.
The LDA densities are indicated by the steps in thick lines. The integer value
of $(\mu(r)+\hbar\Omega)/(2\hbar\Omega)$, (i.e. $I$), is related to the index
$n$ of the intersected Landau level as $I=n+1$. And the relation
$(\mu(r)+\hbar\Omega)/(2\hbar\Omega)=I$ is equivalent to
$\mu(r)= (2n+1)\Omega$.

\noindent Figure.2a: LDA (dotted line) and numerical calculations (solid line)
of the density for $N=2000$ Fermions at $\Omega/\omega_{\perp}$$=0.996$.

\noindent Figure 2b. Density profiles of $N=2000$ Fermions at
$\Omega/\omega_{\perp}$$=0.993$ (crosses), $0.996$ (dots), and 0.998 (solid
line).

\noindent Figure 3a: LDA (dotted line) and numerical calculations (solid line)
of the density for $N=1000$ Fermions at $\Omega/\omega_{\perp}$$=0.982$.

\noindent Figure 3b. Density profiles of $N=2000$ Fermions at
$\Omega/\omega_{\perp}$$=0.982$ (crosses), $0.989$ (dots), and 0.995 (solid
line).

\noindent Figure 4. LDA (dotted line) and numerical calculations (solid line)
of the density for $N=2000$ Fermions at $\Omega/\omega_{\perp}$$=0.99$ and
$\omega_{z}/\omega_{\perp}$$=0.2$.


\begin{references}
\bibitem{width} The extent of erf($x$) is about 1.5, we then have
$\frac{1}{2}\Delta(s/\sqrt{2m_{0}^{\ast}})=1.5$. Since
$\Delta(s) \approx 2r \Delta(r) /a_{\perp}^2$, we have
$\sqrt{m^{\ast}_{0}} \Delta(r) \approx \frac{3}{\sqrt{2}}
\sqrt{m^{\ast}_{0}}a_{\perp}$.
\bibitem{comment} Note that $\rho_{0}(r;L)$
$=\sum_{0}^{L}$$\frac{|w|^{2m}}{m!}$$e^{|w|^2}$.
Since $u_{nm}$$=u_{nm;\lambda=0}$, where
$u_{nm;\lambda}$$=(\pi n!m!)^{-1/2}$
$e^{|w|^2/2}$$\partial_{+}^{m}$$\partial^{n}_{\lambda}$
$e^{\lambda w^{\ast}-|w|^2}$
$=(\pi n!m!)^{-1/2}$
$\partial^{n}_{\lambda}$$[(\lambda-w)^{m}$$e^{\lambda w^{\ast}-|w|^2/2}]$,
(with $a_{\perp}=1$), we have
$\sum_{m=0}^{L}$$|u_{nm;\lambda}|^2$$=\frac{1}{\pi n!}$
$(\nabla^{2}_{\lambda}/4)^{n}$$\sum_{n=0}^{L}$$\frac{|\lambda-w|^{2m}}{m!}$
$e^{\lambda w^{\ast}+\lambda^{\ast}w-|w|^2}$$=\frac{1}{\pi n!}$
$(\nabla^{2}_{\lambda}/4)^{n}$
$[\rho_{0}(\vec{r}-\vec{\lambda}; L)$$e^{\vec{\lambda^2}}]$.
Explicitly,
$\rho_{1}(r; m^{\ast}_{1}) = (1 + \frac{1}{4}\nabla^2)
\rho_{0}(r;m^{\ast}_{1})$,
$\rho_{2}(r;m^{\ast}_{2}) = (1+\frac{1}{2}\nabla^2 + \frac{1}{32} \nabla^4)
\rho_{0}(r;m^{\ast}_{2})$, $\rho_{3}(r;m^{\ast}_{3}) = (1+\frac{3}{4}\nabla^2
+ \frac{3}{32} \nabla^4+\frac{1}{384}\nabla^{6})
\rho_{0}(r;m^{\ast}_{3})$.
\bibitem{temp} The studies of finite temperature effects will be presented
elsewhere. At higher temperatures, the density profile is that of a
Boltzman gas, $\propto e^{-M(\omega^{2}_{\perp}-\Omega^2)r^2/2)/k_{B}T}$.
\end{references}
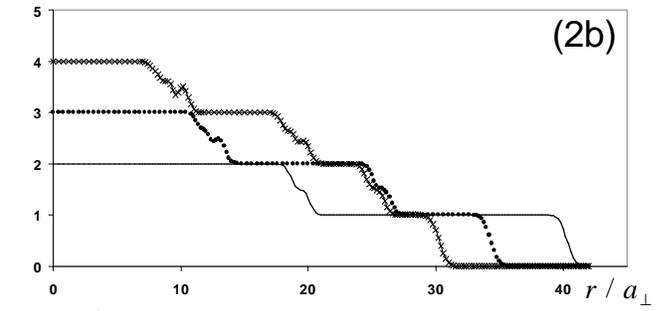
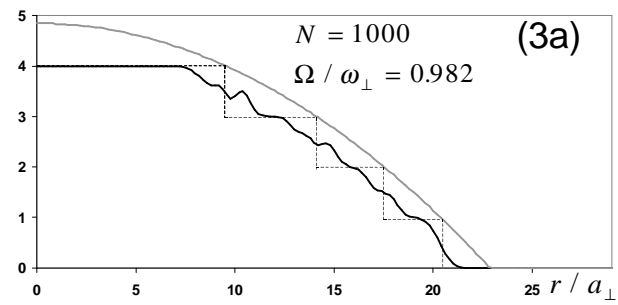
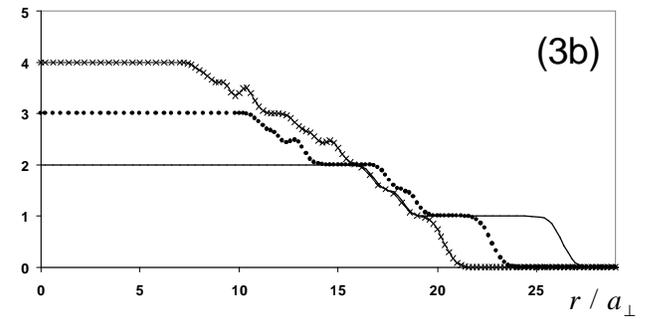
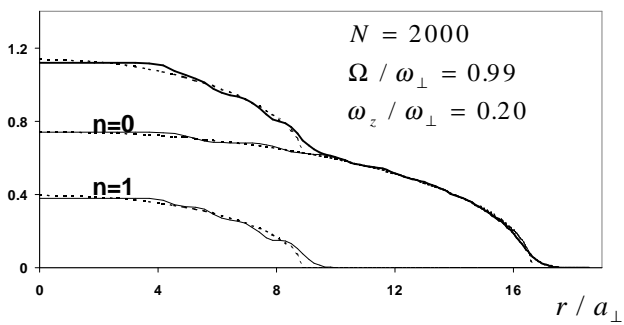
\end{document}